\documentclass[aps,prb,reprint,superscriptaddress]{revtex4-2}

\usepackage[T1]{fontenc}
\usepackage[utf8]{inputenc}
\usepackage{amsmath,amssymb,amsfonts}
\usepackage{bm}
\usepackage{graphicx}
\usepackage{hyperref}
\usepackage{siunitx}
\usepackage{xcolor}
\usepackage{graphicx} 
\usepackage{pgffor}   

\newcommand{\kB}{k_{\mathrm{B}}}

\newcommand{\avgFSi}[1]{\left\langle #1 \right\rangle_{\mathrm{FS},i}}

\begin{document}

\title{Band-Selective Tunneling and Anisotropic Multiband Superconductivity in V$_2$Ga$_5$}
\author{Jozef Haniš}
\email{hanis@saske.sk}
\author{Jozef Kačmarčík}
\affiliation{Centre of Low Temperature Physics, Institute of Experimental Physics, Slovak Academy of Sciences, Ko\v{s}ice, Slovakia}
\author{Filip Košuth}
\affiliation{Centre of Low Temperature Physics, Institute of Experimental Physics, Slovak Academy of Sciences, Ko\v{s}ice, Slovakia}
\author{Levente Faber}
\affiliation{Centre of Low Temperature Physics, Institute of Experimental Physics, Slovak Academy of Sciences, Ko\v{s}ice, Slovakia}
\affiliation{Institute of Physics, Pavol Jozef \v{S}af\'{a}rik University in Ko\v{s}ice, Ko\v{s}ice, Slovakia}
\author{Pavol Szabo}
\affiliation{Centre of Low Temperature Physics, Institute of Experimental Physics, Slovak Academy of Sciences, Ko\v{s}ice, Slovakia}
\author{Szymon Królak}
\affiliation{Faculty of Applied Physics and Mathematics and Advanced Material Center, Gdansk University of Technology, Gda{\' n}sk, Poland}
\author{Micha\l{} J. Winiarski}
\affiliation{Faculty of Applied Physics and Mathematics and Advanced Material Center, Gdansk University of Technology, Gda{\' n}sk, Poland}
\author{Tomasz Klimczuk}
\affiliation{Faculty of Applied Physics and Mathematics and Advanced Material Center, Gdansk University of Technology, Gda{\' n}sk, Poland}
\author{Peter Samuely}
\affiliation{Centre of Low Temperature Physics, Institute of Experimental Physics, Slovak Academy of Sciences, Ko\v{s}ice, Slovakia}
\author{Martin Gmitra}
\email{martin.gmitra@upjs.sk}
\affiliation{Centre of Low Temperature Physics, Institute of Experimental Physics, Slovak Academy of Sciences, Ko\v{s}ice, Slovakia}
\affiliation{Institute of Physics, Pavol Jozef \v{S}af\'{a}rik University in Ko\v{s}ice, Ko\v{s}ice, Slovakia}

\date{\today}

\begin{abstract}
Multiband superconductors with structural anisotropy offer a fertile ground for exploring unconventional quantum states, yet disentangling their directional pairing characteristics remains a formidable challenge. Here, we present a comprehensive thermodynamic and spectroscopic study of the tetragonal intermetallic superconductor $\text{V}_2\text{Ga}_5$ ($T_{\rm c} \approx 3.5$~K), combining first-principles electronic structure calculations with highly sensitive AC calorimetry and directional low-temperature scanning tunneling spectroscopy. By constructing a self-consistent, anisotropic multiband singlet $s$-wave pairing model within the fully symmetric $A_{1g}$ representation, we successfully reconcile the experimental specific heat and upper critical field anomalies. Crucially, we reveal that the apparent reversal of bulk gap hierarchies in directional tunneling experiments is a direct consequence of band-selective tunneling. This effect is governed by an elegant interplay between localized Fermi velocity 'hot spots' and specific Fermi surface topologies, rather than raw thermodynamic gap magnitudes alone. Our findings provide a clear microscopic picture of direction-dependent, band-selective tunneling in a highly uniaxial anisotropic superconductor, demonstrating how orientation-dependent transport constraints shape the observable signatures of multiband quantum condensates.
\end{abstract}

\maketitle

\section{Introduction}
Intermetallic compounds have long served as a fertile playground for exploring complex quantum states, ranging from unconventional magnetism to high-field superconductivity. 
Within the binary V-Ga system, $\text{V}_3\text{Ga}$ (crystallizing in the A15 structure)~\cite{VanVucht1963:PL} historically gained prominence due to its high critical temperature ($T_{\rm c} \approx 15$~K) and widespread utility in high-performance superconducting magnets. 
In contrast, the vanadium-rich gallide $\text{V}_2\text{Ga}_5$ has remained relatively unnoticed, primarily owing to its modest transition temperature of $\approx 3.5$~K~\cite{Kitchingman1972:AC}. 
Recently, however, interest in $\text{V}_2\text{Ga}_5$ has been strongly revitalized by two emerging frontiers in condensed matter physics: the search for topological superconductivity~\cite{Xu2024:PRB} and the discovery of multiband pairing mechanisms in structurally anisotropic systems~\cite{Xu2024:PRB,Huang2025:PRB}.

Structurally, $\text{V}_2\text{Ga}_5$ crystallizes in a tetragonal $P4/mbm$ lattice with the $D_{4h}$ point group (see Fig.~\ref{fig:struct+bands+dos+FS}a). Single crystals typically exhibit a needle-like morphology elongated along the tetragonal $c$-axis~\cite{Teruya2015:JPSJ}, a feature that translates into significantly anisotropic electronic transport properties~\cite{Cruceanu1974:SSC,Huang2025:PRB}. 
On the microscopic scale, the exceptionally short lattice parameter along the $c$-axis ($\approx 2.68$~\AA) facilitates the formation of dense vanadium chains, creating a highly anisotropic electronic environment.

The motivation for investigating $\text{V}_2\text{Ga}_5$ has further intensified following theoretical predictions of a topologically nontrivial normal state~\cite{Xu2024:PRB}, positioning it as a prime candidate for hosting topological superconductivity.
Concurrently, experimental efforts using muon spin relaxation and low-temperature specific heat have provided compelling evidence that its superconducting state cannot be described by a simple, single-gap BCS model~\cite{Lamura2025:SciRep}. 
Instead, these studies report a multi-gap, nodeless $s$-wave behavior, suggesting that multiple Fermi surface sheets contribute distinctly to the superconducting condensate~\cite{Xu2024:PRB}. 
While these macroscopic probes have typically been interpreted through simplified two-gap models to fit the experimental data~\cite{Cheng2024:PRR}, a rigorous microscopic understanding of the individual band contributions and their directional anisotropy has remained missing. 
It turned out that a hidden band-selective transport mechanism dictates quantum signatures of $\text{V}_2\text{Ga}_5$, proving that macroscopic probes can be deeply misleading without a directional microscopic model.

In this work, we resolve this open question by conducting a comprehensive, joint experimental and theoretical investigation of the multiband superconductivity in $\text{V}_2\text{Ga}_5$.
Utilizing first-principles calculations, we establish that the Fermi surface is comprised of three distinct sheets dominated by vanadium $3d$ character, significantly exceeding the gallium $4p$ contributions.
Our calculations reveal a complex interband coupling landscape that necessitates a fully anisotropic, multiband treatment. 
By implementing a fully symmetric singlet $s$-wave pairing model within the $A_{1g}$ representation, we self-consistently fit the pairing matrix to both experimentally obtained specific heat and directional tunneling spectra. The specific heat data were acquired using highly sensitive AC calorimetry on single crystals down to sub-Kelvin temperatures and in high magnetic fields. 
Locally, the superconductivity was probed via low-temperature scanning tunneling microscopy (STM) along both parallel and perpendicular orientations relative to the $c$-axis, yielding reliable surface topographies and directional tunneling conductance ($dI/dV$) spectra.
These directional experiments serve as a crucial benchmark, revealing how the prominent normal-state electronic anisotropy translates directly into a band-selective superconducting pairing state.
Ultimately, our results reconcile recent experimental discrepancies and offer a deeper, unified insight into the pairing symmetry and electronic anisotropy of $\text{V}_2\text{Ga}_5$.

\section{Results}\label{sec:experiment}

\subsection{Electronic heat capacity}\label{sec:heat}
The temperature dependence of the electronic heat capacity contains relevant information about the gap symmetry, the number of bands involved, and the strength of the coupling~\cite{timusk1999pseudogap}.
The normalized heat-capacity jump at the superconducting transition, $\Delta C/\gamma T_{\rm c}$, where $\Delta C = C_{\rm s} - C_{\rm n}$ is the discontinuity between the superconducting and normal-state heat capacities, and $\gamma$ is the Sommerfeld parameter, provides a thermodynamic measure of deviations from the weak-coupling isotropic BCS limit, for which $\Delta C/\gamma T_{\rm c}\approx1.43$. An enhanced jump is commonly associated with strong-coupling superconductivity, whereas a reduced jump can arise from multiband superconductivity, gap anisotropy, or a distribution of superconducting energy scales~\cite{paglione2010high}. Additional information is obtained from the low-temperature dependence, where an exponential suppression of $C(T)$ is expected for a fully gapped superconducting state~\cite{bardeen1957theory}, while a power-law dependence is characteristic of nodal quasiparticle excitations~\cite{Sigrist1991:RMP}.

\begin{figure*}
    \includegraphics[width=0.6\textwidth, angle=0]{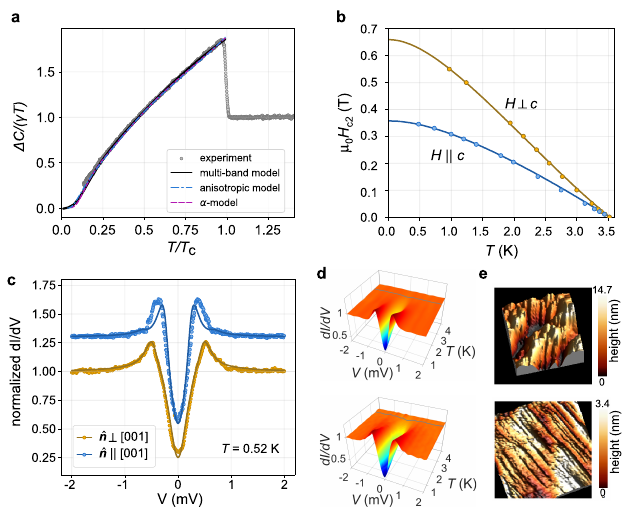}
    \caption{Electronic heat capacity and tunneling characteristics of the V$_2$Ga$_5$.
    \textbf{a}~Measured temperature dependence of the superconducting heat capacity (circles) fitted by the phenomenological $\alpha$-model, anisotropic model, and multigap anisotropic model.
    \textbf{b}~Extracted upper critical magnetic fields at half the transition as a function of temperature for $H\parallel c$ (blue circles) and for $H\perp c$ (yellow circles). The solid lines are a fit of the Gurevich two-band model.
    \textbf{c}~Typical tunneling spectra measured at $T = 0.52$~K with tunneling current parallel (blue symbols) and perpendicular (yellow symbols) to the $c$ axis. The solid lines are directional tunneling conductance calculated using a multiband anisotropic model
    with $\Gamma = 10~\mu$eV for ($\hat{\bm{n}}\parallel [001]$), and with $\Gamma = 20~\mu$eV for averaged in-plane orientations ($\hat{\bm{n}}\perp [001]$). In both cases, the directionality parameter was fixed to $\theta =0.01$. 
    An offset for the $\hat{\bm{n}}\parallel [001]$ case was used for clarity.
    \textbf{d}~Temperature dependent tunneling spectra for tunneling current parallel to $c$-axis (top) and perpendicular to the $c$-axis (bottom). The gray spectra correspond to the bulk critical temperature $T_{\rm c} = 3.5$~K.
    \textbf{e}~Surface topography measured in direction parallel to $c$ axis on $200 \times 200$~nm$^2$ (upper) and the direction perpendicular to $c$ axis on $400 \times 400$~nm$^2$ (bottom).
    }
\label{fig:Cv_Hc2_experiment}
\end{figure*}

In Figure~\ref{fig:Cv_Hc2_experiment}a, we show the temperature dependence of the measured normalized electronic heat capacity $\Delta C/\gamma T_{\rm c}$.
The resulting transition at $T_{\rm c}$ exhibits a sharp discontinuity characteristic of a second-order phase transition. 
The observed jump height, $\Delta C/\gamma T_{\rm c} \approx 0.8$, is significantly reduced compared to the isotropic BCS limit. 
Also, the overall temperature dependence is very different from the BCS prediction.
A narrow superconducting transition of the second order characterizes the quality of the sample. The thermodynamic superconducting transition temperature at zero field was determined from the local entropy balance around the phase transition, giving $T_{\rm c} = 3.5$~K.

Magnetic-field-dependent measurements show significant anisotropy appearing in the superconducting field response. 
The superconducting anomaly is progressively suppressed by the magnetic field for both field orientations,  parallel and perpendicular to the crystallographic $c$-axis, and the anisotropic upper critical fields at zero temperature $H_{c2\parallel c}(0)\approx 0.37$~T and $H_{c2\perp c}(0)\approx 0.66$~T were extracted (see Supplementary Note 1).
In Fig.~\ref{fig:Cv_Hc2_experiment}b, we show the upper critical fields as a function of temperature for both magnetic field orientations.
The solid lines are theoretical predictions based on the Gurevich model \cite{Gurevich2003:PRB}, assuming intraband and interband couplings as discussed in the next section.

Temperature dependence of the heat capacity is routinely quantified within the framework of the phenomenological $\alpha$-model.
In this case, the total electronic specific heat, $C$, is treated as a weighted sum of independent BCS-like contributions: $C(T) = \sum_i \gamma_i C^{\rm BCS}(\Delta_i, T)$, where the weights $\gamma_i$ satisfy the condition $\sum_i \gamma_i = \gamma_{\rm n}$. 
A central feature of this $\alpha$-model is that the coupling ratios $2\Delta_{i,0}/k_{\rm B}T_{\rm c}$ are not fixed to the weak-coupling BCS value of 3.53, but are instead treated as free parameters. 
For a two-gap system, this approach reduces to the Bouquet model \cite{Bouquet2001:EPL}, requiring only three independent fitting parameters: $\Delta_1$, $\Delta_2$, and the relative weight $w = \gamma_1/\gamma_{\rm n}$.
The measured data show a negative curvature below $T_{\rm c}$ down to the lowest measured temperature, indicating the presence of smaller gap(s) in the superconducting quasiparticle spectrum. 
The temperature dependence can be fitted by the alpha model with two gaps $\Delta_{1}/k_{\rm B}T_{\rm c} = 1.8$ and $\Delta_{2}/k_{\rm B}T_{\rm c} = 0.65$ with the weights $w = \gamma_1/\gamma_{\rm n} = 0.55$ and $w = \gamma_2/\gamma_{\rm n} = 0.45$, see dashed line in Fig.~\ref{fig:Cv_Hc2_experiment}a.

We note that the measured data can also be described by an effective single-band anisotropic $s$-wave model, where the gap anisotropy is supposed to be in the form $\Delta = \Delta_0 [1 + \delta \cos (4\phi)]$ accounting for the tetragonal in-plane anisotropy of V$_2$Ga$_5$ with strength given by the $\delta$ parameter. 
The fit shown in Fig.~\ref{fig:Cv_Hc2_experiment}a by dash-dotted line for $\delta = 0.68$ and $2\Delta_0/k_{\rm B}T_{\rm c} = 2.5$ parameters.
The overall difference in the temperature dependence of $\Delta C$ between the two-gap and anisotropic gap models is negligible.

Despite their computational simplicity and ability to provide high-quality fits to experimental data, the $\alpha$-model and anisotropic gap model remain purely phenomenological. 
Assumption of non-interacting bands is often physically ungrounded, as it can yield parameters that are inconsistent with other thermodynamic observables, such as the upper critical field $H_{\rm c2}$ \cite{Xu2024:PRB}.
Additionally, a weak interband coupling imposes the same $T_{\rm c}$ \cite{Suhl1959:PRL,Tang1970:PRB} in multigap systems.
Furthermore, approximation of unique gap anisotropy may fail to disclose the true momentum-dependent physics of the system reflected in directional tunneling conductance measurements. 

\subsection{STM measurements}
Surface topography measurements were carried out in constant-current mode, using bias voltages in the range of $10 - 100$~mV and tunneling currents of $0.5 - 1$~nA. 
Representative surface topographies are presented in Fig.~\ref{fig:Cv_Hc2_experiment}e.
The scan acquired parallel to the $c$-axis exhibits a typical needle-like morphology.
Correspondingly, the surface perpendicular to the $c$-axis consists of periodically repeating parallel terraces with widths of approximately $20 - 40$~nm, aligned along a single crystallographic direction, which indicates the presence of pronounced twinning.

The magnitude and possible anisotropy of the superconducting energy gap were investigated by means of local STM tunneling spectroscopy. 
Several thousand \textit{I–V} characteristics were recorded on cleaved surfaces in different crystallographic orientations, using tunneling junctions with a typical resistance of approximately 10 $\text{M}\Omega$. 
The corresponding differential conductance spectra (\textit{dI/dV}) were obtained by numerical differentiation of the measured \textit{I–V} curves. 
Typical tunneling spectra acquired along and perpendicular to the $c$-axis are shown in Fig.~\ref{fig:Cv_Hc2_experiment}c.
Both spectra indicate superconducting behavior with pronounced symmetric coherence peaks and a V-shaped conductance profile within the gap.
Notably, the peak positions differ between the two orientations: the peak-to-peak voltage is about 0.7~mV for measurements performed parallel to the $c$-axis, and about 1~mV for those measured perpendicular to the $c$-axis. 
This demonstrates the directional dependence of the superconducting energy gap.
In Fig.~\ref{fig:Cv_Hc2_experiment}d, we present the 3D perspective waterfall plots of the temperature-dependent tunneling spectra. 
As temperature increases, the superconducting features in both tunneling geometries are gradually suppressed, and the transition to the normal state occurs at $T_{\rm c} \approx 3.5$~K in both cases (represented by the gray spectra), consistent with the bulk transition temperature observed in heat capacity, see Fig.~\ref{fig:Cv_Hc2_experiment}a.
This agreement argues against an unrelated surface superconducting phase as an origin of the directional spectra and indicates that STM probes the same superconducting state detected by bulk thermodynamics.
We note that the temperature dependencies are well described by the proposed multiband superconducting anisotropic model, see Supplementary Note 2.

Magnetic-field-dependent tunneling spectra measured at $T = 0.5$~K for both crystallographic directions are provided in Supplementary Note 1.
The superconducting gap closes at upper critical fields of $H_{{\rm c}2||c}\approx 0.37$~T for fields applied parallel to the $c$-axis and $H_{{\rm c}2 \perp c}\approx 0.7$~T for fields applied perpendicular to the $c$-axis. The close agreement between the values of $T_{\rm c}$, $H_{{\rm c}2||c}$ and $H_{{\rm c}2 \perp c}$ obtained from STM and heat-capacity measurements indicates representative bulk superconductivity.

Our tunneling spectroscopy data consistently reveal a V-shaped gap with orientation-dependent gap magnitude and upper critical field, while the critical temperature remains isotropic.
The line shape of the spectra with an additional weak shoulder-like feature at 0.25~mV for the tunneling current perpendicular to the $c$-axis provides a second indication that the superconducting state can not be described by an isotropic single-gap superconductor.
These spectral features motivate the multiband anisotropic modelling that accounts for both the thermodynamic and tunnelling responses. 

\subsection{Electronic structure}
\begin{figure*}[ht]
    \includegraphics[width=0.98\textwidth]{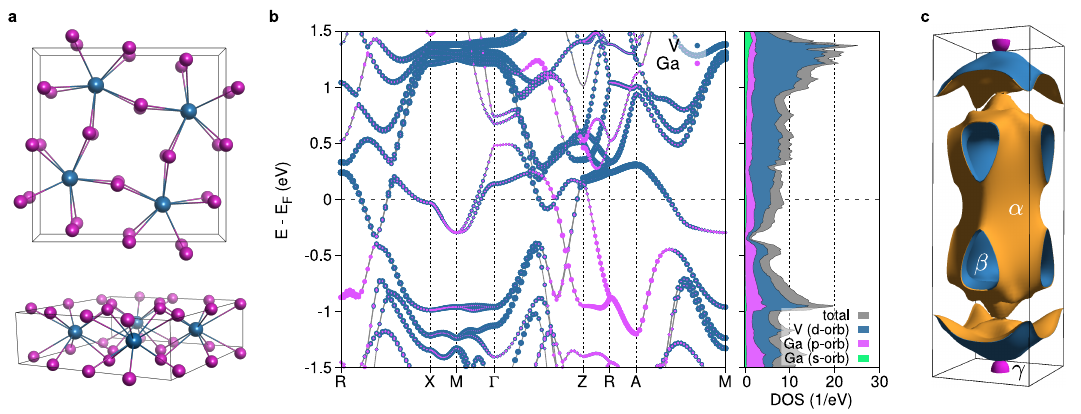}
    \caption{Atomic structure of V$_2$Ga$_5$ crystal and calculated electronic structure.
    \textbf{a}~Top and perspective view, V atoms are blue and Ga atoms magenta.
    \textbf{b}~Electronic bands along high-symmetry lines with projected states on Ga and V atoms, and total and projected density of states on V $d$-orbitals and Ga $p$ and $s$-orbitals.
    \textbf{c}~Band index resolved Fermi surface with $\alpha$ (ochre), $\beta$ (blue), and $\gamma$ (magenta) bands.
    }
    \label{fig:struct+bands+dos+FS}
\end{figure*}

Figure~\ref{fig:struct+bands+dos+FS}b shows the calculated electronic bands along lines connecting high-symmetry points in the first Brillouin zone.
The electronic states are projected onto the V $3d$ orbitals and Ga $4p$ orbitals.
The bands corresponding to V $3d$ orbitals disperse with a significant $k_z$ component, while the bands for Ga $4p$ orbitals dominate at in-plane momenta.
The calculated density of states is dominated by V $3d$ states, which are approximately twice as prominent as the Ga $4p$ contributions. 
This electronic dominance, coupled with the remarkably short V--V interatomic distance along the $c$-axis, supports the description of V$_2$Ga$_5$ as a quasi-1D superconductor \cite{Xu2024:PRB,Huang2025:PRB,Lamura2025:SciRep}.
In this framework, superconductivity has been expected to be primarily hosted by the vanadium chains, while the gallium sublattice serves to electronically decouple these chains, leading to the observed multiband characteristics and anisotropic critical magnetic fields.
However, the directional tunneling conductance parallel to the $c$-axis is affected by the Ga $p$-orbitals.

In Fig.~\ref{fig:struct+bands+dos+FS}c, we plot band-resolved Fermi surfaces of V$_2$Ga$_5$ in the first Brillouin zone.
The Fermi surface comprises two $\Gamma$-centered hole-corrugated quasi-two-dimensional cylindrical pockets with multiple necking and wavy caps along $k_z$ (the $\alpha$ and $\beta$ sheets), and a small Z-centered ellipsoidal pocket (the $\gamma$ sheet). 
Projection of the V $d$-orbitals on the Fermi surface populates states mostly on the wavy parts of the $\alpha$ and $\beta$ sheets with large in-plane momenta and $|k_z|\approx 0.77\pi/c$, and the capping of $\Gamma$-centered cylindrical pockets of the $\beta$ sheet.
The Ga $p$-orbitals predominantly populate the capping of the $\Gamma$-centered cylindrical $\alpha$ pockets and the $\gamma$ pocket, see Supplementary Note 3.

To support our DFT electronic structure, we calculated the de Haas-van Alphen frequencies using the \textsc{skeaf} code \cite{Rourke2012:CPC} and identified the lowest frequency of about 126~T, which corresponds to the extremal cross-section of the $\gamma$ pocket and is in good agreement with the recent measurements \cite{Huang2025:PRB}.

\subsection{Three-band anisotropic superconducting model}
The superconducting phase is described using a three-overlapping-band model and fully symmetric singlet $A_{1g}$ pairing \cite{Sigrist1991:RMP} in the $D_{4h}$ point group, for more details, see the Method section.
The momentum-dependent pairing function in this case contains uniaxial anisotropy along the $c$-axis and in-plane (perpendicular to the $c$-axis) four-fold anisotropy controlled by the $a$ and $b$ parameters, respectively.
To determine the band coupling matrix elements $V_{ij}$ and the anisotropy parameters $a$ and $b$ entering the pairing basis, we solve the self-consistent gap equations and fit the resulting temperature dependencies of the specific heat and directional tunneling conductances to the experimental data. 
For the fitting procedure, we employ hyperparameter optimisation using Optuna \cite{akiba2019optuna}, by minimizing a least-squares objective function. 
In the final parameter set, we obtain $a=0.204$, $b=0.180$, and
\begin{equation}
V_{ij} = \left(
\begin{array}{ccc}
 93.11 & 22.05 & 3.80 \\
 22.05 & 35.38 & 23.84 \\
 3.80 &  23.84 & 0.13
\end{array}\right).
\label{eq:lambda_res}
\end{equation}
Obtaining positive values for the parameters $a$ and $b$ implies fully gapped, nodeless superconductivity. A ratio $a/b > 1$ enhances the gap magnitude for momenta closer to the $c$-axis while maintaining a fourfold in-plane gap modulation (see inset to Fig.~\ref{fig:gap_FS_dIdV}a). 
In Fig.~\ref{fig:Cv_Hc2_experiment}a, we compare the temperature dependence of the calculated specific heat to the measured data. Figure~\ref{fig:gap_FS_dIdV}a shows the reduced temperature dependencies of the self-consistently calculated gaps $\Delta_i/k_{\rm B}T_{\rm c}$ for $i=\{\alpha,\beta,\gamma\}$, with $T_{\rm c}=3.5$~K. 
Furthermore, Fig.~\ref{fig:gap_FS_dIdV}b illustrates the color-coded gap amplitude projected onto the Fermi surface sheets, including cross sections through the zone center along the (100) and (001) planes.
The calculated normalized gaps are compared with the $\alpha$-model in Table~\ref{tab:Deltas}.
\begin{table}[h!]
    \centering
    \setlength{\tabcolsep}{10pt}
    \begin{tabular}{c||cc}
    \hline
    $\Delta_i(T\to 0)/k_{\rm B}T_{\rm c}$ & $\alpha$-model & our work \\ \hline\hline
    gap $\alpha$ & $1.8$  & $1.87$\\
    gap $\beta$ & $0.65$ & $0.66$\\
    gap $\gamma$ & not resolved & $0.24$ \\
    \hline
    \end{tabular}
    \caption{Comparison of normalized gaps $\Delta_i/k_{\rm B}T_{\rm c}$ at $T\to 0$ from experiment and our multiband approach.
    The $\alpha$-model fitted to the data with relative weight $w=0.55$ for the larger gap.
    }
    \label{tab:Deltas}
\end{table}

The fitted interaction matrix is heavily dominated by the $\alpha$ Fermi sheet, yielding an intraband coupling of $V_{\alpha\alpha}=93.11$~meV alongside a substantial interband term of $V_{\alpha\beta}=22.05$~meV. 
In comparison, the intraband pairing strength on the $\beta$ sheet ($V_{\beta\beta}=35.38$~meV) is significantly weaker, while the coupling on the $\gamma$ sheet is virtually negligible ($V_{\gamma\gamma}=0.13$~meV), suppressed by nearly three orders of magnitude relative to $V_{\alpha\alpha}$. 
Despite this pronounced disparity, the interband coupling $V_{\alpha\gamma}$ remains sizable and effectively induces superconductivity within the passive $\gamma$ pocket -- a feature that becomes highly relevant when interpreting directional tunneling configurations. 
This hierarchy of diagonal elements underscores that the pairing condensate is strongly anchored on the $\alpha$ sheet, specifically within its wavy segments at $|k_z|\approx 0.77\pi/c$ and at the caps of the $\Gamma$-centered cylindrical pocket near $|k_z|\approx 0.68\pi/c$ (see Fig.~\ref{fig:gap_FS_dIdV}b). 
Crucially, however, our subsequent analysis of the directional tunneling conductance reveals a striking contrast: the subdominant $\beta$ band completely dominates the transport signature along the $c$-axis, whereas the intrinsically stronger $\alpha$ band governs tunneling perpendicular to the $c$-axis, albeit with an effectively reduced amplitude dictated by its anisotropic pairing ratio $b<a$.

\subsection{Upper-critical-field analysis}
The distribution of pairing strengths, $V_{\alpha\alpha} > V_{\beta\beta} \gg V_{\gamma\gamma}$, suggests that an effective two-band model accounting for the $\alpha$ and $\beta$ bands can capture the primary superconducting response at the phenomenological level (see Table~\ref{tab:Deltas} and Refs.~\cite{Cheng2024:PRR,Lamura2025:SciRep}). 
This simplification is further justified by the small reduced gap $\Delta_{\gamma}/k_B T_c \approx 0.24$, which indicates that the $\gamma$ band is only weakly coupled to the superconducting condensate. 
In the following, we describe the temperature dependence of the upper critical field using the Gurevich two-band model~\cite{Gurevich2003:PRB}. 
The equation for $H_{\rm c2}(T)$ is defined as:
$a_{0}[{\rm ln}(t)+U(h)][{\rm ln}(t)+U(\eta h)] + a_{1}[{\rm ln}(t) + U(h)] + a_{2}[{\rm ln}(t) + U(\eta h)] = 0$, where $a_{0} = 2(\lambda_{11} \lambda_{22} - \lambda_{12} \lambda_{21})$, $a_{1/2} = 1 \pm (\lambda_{11} - \lambda_{22})/\lambda_{0}$, and $\lambda_{0} = \sqrt{(\lambda_{11} - \lambda_{22})^2 + 4 \lambda_{12} \lambda_{21}}$. 
Here, $h = H_{\rm c2}D_{1}/(2\phi_{0}T)$, $t = T/T_{\rm c}$, $\eta = D_2/D_1$, and $U(x) = \psi(x+1/2) - \psi(1/2)$, where $\psi(x)$ is the digamma function and $\phi_0$ is the flux quantum. 
The coupling constants $\lambda_{ij}$ were taken as $\lambda_{ij} = N_j(0)V_{ij}$, with the partial densities of states $N_j(0)$ obtained from DFT calculations at the Fermi level (see Supplementary Note 3) and coupling matrix elements as specified by Eq.~(\ref{eq:lambda_res}). 
The band diffusivities $D_1$ and $D_2$ were treated as fitting parameters, yielding $\eta_{H\perp c}=9.65$ and $\eta_{H\parallel c}=0.8$. 
The fitted diffusivity ratios reflect pronounced band- and orientation-dependent transport anisotropy. 
The large value of $\eta_{H\perp c}$ indicates a significant disparity in the transport properties of the $\alpha$ and $\beta$ bands when quasiparticles are driven along the $c$-axis.
Conversely, for $H \parallel c$, the diffusivities become comparable.
The resulting temperature dependencies of the $H_{\rm c2}(T)$ for fields applied parallel and perpendicular to the $c$-axis are shown as solid lines in Fig.~\ref{fig:Cv_Hc2_experiment}b.

\subsection{Tunneling conductance calculations}
The tunneling density of states is evaluated using the Dynes formalism \cite{Dynes1978:PRL,Herman2016:PRB}, with separate anisotropic gaps and finite lifetime broadening parameters.
In the multigap V$_2$Ga$_5$ system, the total differential conductance is a convolution of velocity-weighted densities of states and band-specific tunneling matrix elements $|T_{i\bm{k}}|^2=v_{\rm g} D(\bm{k})$, with the directionality function $D(\bm{k})=\exp(-\frac{\bm{k}^2-(\bm{k}\cdot\hat{\bm{n}})^2}{(\bm{k}\cdot\hat{\bm{n}})^2 \theta^2})$ \cite{Ledvij1995:PRB,Yusof1998:PRB}, where $\theta$ is an angular tip spread parameter and $\hat{\bm{n}}$ is the tunneling directional unit vector.

\begin{figure*}
  \centering
  \includegraphics[width=0.6\linewidth]{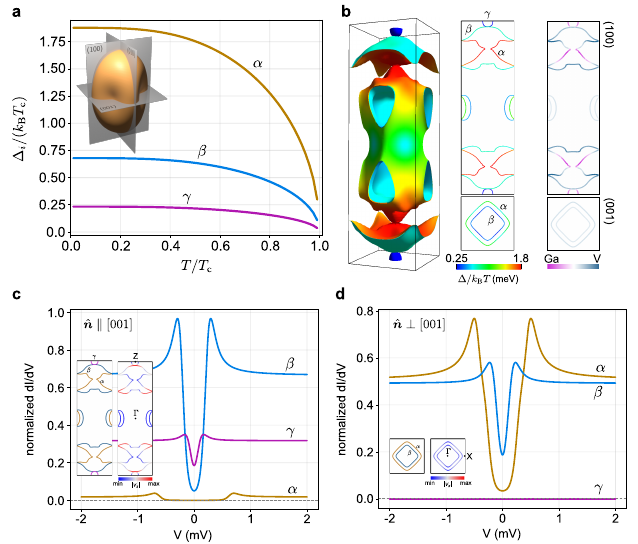}
  \caption{Calculated anisotropic multiband superconducting characteristics of V$_2$Ga$_5$.
  \textbf{a}~Temperature dependencies of the band-resolved gap amplitudes. The inset shows the angular part of the singlet fully symmetric $A_{1g}$ anisotropic pairing given by Eq.~ (\ref{eq:phi_sph}) with $a=0.204$ and $b=0.180$ and principal planes centered at the Brillouin zone center.
  \textbf{b}~Superconducting gap amplitudes plotted on Fermi surface sheets and section plots along (100) and (001) surfaces with corresponding Fermi contour plots. The contour plots with the atomic orbital Ga $p$ and V $d$ character are also shown.
  \textbf{c}~Band resolved differential conductance for $\hat{\bm{n}}\parallel [001]$ with $\Gamma = 10~\mu$eV. Insets show (100) section plane of the Brillouin zone with band-resolved Fermi contours and $|v_z|$ amplitudes for the states at the contours.
  \textbf{d}~Band resolved differenctial conductance for $\hat{\bm{n}}\perp [001]$ with $\Gamma = 20~\mu$eV. Insets show the (001) section plane of the Brillouin zone with band-resolved Fermi contours and $|v_x|$ amplitude for the states at the contours.
    }
  \label{fig:gap_FS_dIdV}
\end{figure*}

In Fig.~\ref{fig:Cv_Hc2_experiment}c, we plot the calculated normalized differential conductance for $\theta = 0.01$ at $T=0.52$~K considering two tunneling orientations: a surface normal parallel to the crystallographic $c$-axis ($\hat{\bm{n}}\parallel [001]$) and an in-plane surface orientation ($\hat{\bm{n}}\perp [001]$). For $\hat{\bm{n}}\parallel [001]$, the calculated spectrum exhibits coherence peaks at a bias of $\pm 0.31$~mV. For the in-plane case ($\hat{\bm{n}}\perp [001]$), we average the calculated $dI/dV$ spectra over several in-plane directions to mimic the experimental uncertainty in the surface orientation and termination, yielding peaks at $\pm 0.51$~mV. While the resulting spectra match the experimental data remarkably well (Fig.~\ref{fig:Cv_Hc2_experiment}c), a direct evaluation of the peak-to-peak distances suggests that the gap in the in-plane direction dominates the gap along the $c$-axis. Crucially, this observation stands in stark contrast to our calculated bulk anisotropic gap structure, where $a > b$ indicates that the intrinsic gap amplitude is actually maximal along the $c$-axis (see inset to Fig.~\ref{fig:gap_FS_dIdV}a).

To resolve this apparent contradiction, we investigate the orientation-dependent band selectivity of the tunneling process. In Figs.~\ref{fig:gap_FS_dIdV}c and \ref{fig:gap_FS_dIdV}d, we isolate the calculated $dI/dV$ spectra for each band. For tunneling parallel to the $c$-axis ($\hat{\bm{n}}\parallel [001]$), the transport signature is not dominated by the intrinsically larger $\alpha$ band, but rather by the subdominant $\beta$ band, which features a prominent coherence peak at $\pm 0.31$~mV alongside a minor low-bias shoulder from the $\gamma$ pocket. Conversely, transitioning to the perpendicular orientation ($\hat{\bm{n}}\perp [001]$) induces a complete redistribution of the spectral weights: the $\alpha$ band becomes dominant, exhibiting its main coherence peaks at $\pm 0.51$~mV, while the $\gamma$ pocket contribution vanishes entirely. This striking redistribution highlights the behavior of a highly uniaxial anisotropic system where momentum-dependent tunneling matrix elements, rather than raw thermodynamic gap magnitudes alone, dictate the experimental observables.

This band-selective tunneling mechanism is microscopically illustrated by the Brillouin zone cross-sections presented as insets to the respective transport directions. For transport parallel to the $c$-axis, the momentum distribution of the velocity component $|v_z|$ (inset to Fig.~\ref{fig:gap_FS_dIdV}c) reveals distinct 'hot spots' of maximal velocity (red regions) concentrated exclusively on the $\beta$ Fermi surface sheet near the zone boundary. Because the transmission weight scales with $|\mathbf{v}_i \cdot \hat{\bm{n}}|$, these high-velocity states strongly channel the tunneling current into the $\beta$ band, leaving the $\alpha$ band spectroscopically muted despite its larger thermodynamic gap. Conversely, for perpendicular tunneling within the (001) plane (inset to Fig.~\ref{fig:gap_FS_dIdV}d), the in-plane velocity component $|v_x|$ is virtually uniform across both the $\alpha$ and $\beta$ sheets. In the absence of a velocity-driven selection advantage, the tunneling contribution is instead governed by the phase-space area of the Fermi surface contours. Consequently, the large-gap $\alpha$ band reclaims dominance in the perpendicular spectrum simply due to its larger Fermi wavevector $k_{\rm F}$ and expanded contour geometry, which offer a significantly higher density of available tunneling channels.
The $\beta$ band contributes as the weak, shoulder-like feature observed at approximately $\pm 0.25$~mV in the perpendicular tunneling conductance.

\section{Discussions}\label{sec:conclusions}
In summary, we have developed a self-consistent, three-band anisotropic singlet $s$-wave model within the fully symmetric $A_{1g}$ representation to describe the superconducting state of V$_2$Ga$_5$. 
By scaling the pairing amplitudes and angular anisotropy parameters of the superconducting gap function to fit the experimental specific heat and directional tunneling conductances, we show that the superconducting gap features a pronounced uniaxial anisotropy along the $c$-axis paired with a fourfold in-plane modulation.
Crucially, the model resolves how band-selective tunneling driven by Fermi velocity 'hot spots' projections and specific Fermi surface topologies can suppress the apparent coherence peak positions in raw experimental $dI/dV$ spectra. 
While the robust intraband pairing potentials of the $\alpha$ and $\beta$ bands dominate the pairing matrix, the weak intraband strength of the $\gamma$ pocket confirms that its superconductivity near the $Z$-point is entirely induced via interband cross-coupling. 
The passive nature of the $\gamma$ pocket is independently validated by the success of a simplified, effective two-band model in capturing the upper critical field $H_{{\rm c}2}(T)$ profiles from specific heat thermodynamics. 
Together, these convergent thermodynamic and spectroscopic insights highlight the vital interplay between structural uniaxial anisotropy and interband coupling in dictating the macroscopic superconducting properties of V$_2$Ga$_5$ quantum condensate.

\section{Methods}
\subsection*{Sample preparation and heat capacity setup}
Single crystals of V$_2$Ga$_5$ were grown in Al$_2$O$_3$ crucibles using V powder (Alfa Aesar, 99.5\%) and Ga pieces (Onyxmet, 99.99\%), with a 6:94 V:Ga molar ratio.
The evacuated quartz tube was heated to $1000^\circ\mathrm{C}$ at a rate of $100^\circ\mathrm{C}\,\mathrm{h}^{-1}$, held at this temperature for 48~h, and subsequently cooled to $550^\circ\mathrm{C}$ at a rate of $2.5^\circ\mathrm{C}\,\mathrm{h}^{-1}$. At this temperature, the tubes were centrifuged to remove excess Ga flux. The resulting needle-like crystals, with a typical length of 3 mm, were etched in lightly diluted hydrochloric acid (HCl) to eliminate residual Ga from the surface, following the procedure described in Ref.~\cite{Cheng2024:PRR}.
V2Ga5 crystals were found to be stable in air.

The ac heat capacity measurements in a magnetic field up to 8~T are performed by a calorimeter consisting of a chromel constantan thermocouple. The thermocouple served at the same time as a thermometer. The heat was supplied to the sample from an LED through an optical fiber. Such a configuration enables high-resolution measurement of the sample’s heat capacity, however, only in arbitrary units. The measurements were performed in a horizontal superconducting magnet, allowing for the positioning of the crystal with respect to the magnetic field direction. More details are described elsewhere \cite{Volavka2026:PRL}.

\subsection*{STM setup}
The STM experiments were carried out on a home-made STM head attached to the cold plate of a Janis $^3$He cryomagnetic system, which allows measurements at low temperatures down to 0.4 K and in magnetic fields up to 8 Tesla. 
The STM current was measured using an Au tip, which was fabricated in situ by controlled collision of the tip with a massive piece of Au at low temperatures. 
In our experiments, we measured two V$_2$Ga$_5$ samples during a single cooling cycle; these were glued with Ag paint to a grounded copper substrate along with a solid piece of Au. 
The surfaces of the needle-shaped samples were prepared by cleaving in two crystallographic directions. 
The sample with a surface perpendicular to the $c$-axis was prepared by crosscutting needles with a sharp steel blade.
The second sample, with a surface parallel to the $c$-axis, was broken with a small hammer in a special cleaver. Before being placed in the vacuum chamber of the STM system, the samples were exposed to ambient conditions for approximately 20 minutes.

\subsection*{Electronic structure calculations}
For electronic structure calculation, we used the \textsc{Quantum Espresso} code \cite{Giannozzi2017} with lattice constant $a=8.9716$~\AA\ and $c=2.69272$~\AA\ \cite{Cheng2024:PRR}.
Atomic positions within the cell were relaxed using the BFGS quasi-Newton algorithm \cite{Fletcher1987}. 
We used projector augmented-wave pseudopotentials \cite{Bloechl1994:PRB} \verb|V.rel-pbe-spnl-kjpaw_psl.1.0.0.UPF| and \verb|Ga.rel-pbe-dnl-kjpaw_psl.1.0.0.UPF| from pslibrary \cite{DalCorso2014CMS} including spin-orbit coupling and generalized gradien approximation for exchange-correlation functional \cite{Perdew1996:PRL}.
For the kinetic energy cutoff for wavefunctions and charge density, 50~Ry and 650~Ry were considered.
For Brillouin zone sampling, we used $6\times 6 \times 20$ $k$-points in the irreducible wedge for self-consistent calculations.
The Fermi surface states involved in the solution of the gap equations were sampled with $24\times 24\times 80$ $k$-points.

\subsection*{Anisotropic three-band pairing model}\label{sec:model}
We follow the multiband extension of the overlapping-band model \cite{bardeen1957theory, Suhl1959:PRL}, considering fully symmetric singlet $A_{1g}$ pairing \cite{Sigrist1991:RMP} for the $D_{4h}$ point group.
The multiband superconductivity can be described by the BCS pairing Hamiltonian
\begin{align}
H &= \sum_{i,\bm{k},\sigma} \xi_{i\bm{k}}\, c^{\dagger}_{i\bm{k}\sigma} c_{i\bm{k}\sigma}
-\sum_{i,j}\sum_{\bm{k},\bm{k}'}
V_{ij,\bm{k}\bm{k}'}
c^{\dagger}_{i\bm{k}\uparrow}c^{\dagger}_{i-\bm{k}\downarrow}
c_{j-\bm{k}'\downarrow}c_{j\bm{k}'\uparrow},
\label{eq:H}
\end{align}
where $V_{ij,\bm{k}\bm{k}'}$ couples a singlet Cooper pair from band $j$ to band $i$, and $\xi_{i\bm{k}}=\varepsilon_{i\bm{k}}-\mu$, is the single-particle energy dispersion with respect to the chemical potential.
At the mean-field level, the quasiparticle energy equals
\begin{equation}
E_{i\bm{k}}(T)=\sqrt{\xi_{i\bm{k}}^{2}+\Delta^2_{i\bm{k}}(T)},
\end{equation}
where $\Delta_{i\bm{k}}(T)$ is the band-resolved superconducting order parameter.
The self-consistency gap equation for $T<T_{\rm c}$ reads \cite{Mineev1999}
\begin{equation}
\Delta_{i\bm{k}}(T)=
-\sum_{j}\sum_{\bm{k}'}
V_{ij,\bm{k}\bm{k}'}
\frac{\Delta_{j\bm{k}'}(T)}{2E_{j\bm{k}'}(T)}
\tanh\!\left(\frac{E_{j\bm{k}'}(T)}{2k_{\mathrm B}T}\right).
\label{eq:gap_full}
\end{equation}
In the weak-coupling treatment, only states below the cutoff energy $|\xi_{j\bm{k}}|\le \hbar\omega_c$ contribute to pairing.
To obtain a tractable anisotropic model, we assume a separable pairing kernel \cite{Markowitz1963:PR},
\begin{equation}
V_{ij,\bm{k}\bm{k}'}=-V_{ij}\,\phi_i(\bm{k})\,\phi_j(\bm{k}'),
\label{eq:V_stheep}
\end{equation}
where $V_{ij}$ is the symmetric coupling matrix.
Equation~\eqref{eq:gap_full} then admits a separable gap on the $i$th Fermi-surface sheet,
\begin{equation}
\Delta_{i\bm{k}}(T)=\Delta_i(T)\,\phi_i(\bm{k}),
\label{eq:Delta_sep}
\end{equation}
where $\phi_i(\bm{k})$ is the normalized anisotropy part of the pairing function
$\phi_i(\bm{k})=\tilde{\phi}_i(\bm{k}) / \sqrt{\langle \tilde{\phi}_i^2\rangle_{\mathrm{FS}_i}}$,
where $\langle\ldots\rangle_{{\rm FS}_j}$ denotes the Fermi-surface average over the $i$-th band.
To get the anisotropy of the gap that depended only on direction and not on radial distance from the Fermi surface, we used $\bm k = \bm{k}_{\rm BZ} / |\bm{k}_{\rm BZ}|^2$.
For the angular part of the pairing, we consider a fully symmetric function corresponding to the $A_{1g}$ representation of the $D_{4h}$ point group, which reads
\begin{equation}
\tilde{\phi}_j(\bm{k}) = 1 + a f_2(\bm{k}) + b f_4(\bm{k})
\label{eq:phi_sph}
\end{equation}
with the isotropic term, the uniaxial anisotropy term $f_2(\hat{\mathbf{k}})=2\hat{k}_z^2-(\hat{k}_x^2+\hat{k}_y^2)$, and the fourfold in-plane anisotropy term $f_4(\hat{\mathbf{k}})=\hat{k}_x^4+\hat{k}_y^4-6\hat{k}_x^2\hat{k}_y^2$.
The dimensionless parameters $a$ and $b$ control the anisotropy of the gap, and they were determined by the procedure involving self-consistent treatment of the gap equations~\eqref{eq:gap_full}, calculating specific heat and fitting its temperature dependence to the measured data.

\subsection*{Electronic heat capacity}
The electronic heat capacity follows from differentiating the entropy, $C(T)=T dS(T)/dT$, which is given by
\begin{equation}
  S(T) =
  -2\kB \sum_i N_i(0)
  \avgFSi{
    \int_{-\infty}^{\infty} dE 
    {\cal F}_i
  },
  \label{eq:entropy}
\end{equation}
where ${\cal F}_i=\left[ f_{i}\ln f_{i} + (1-f_{i})\ln(1-f_{i}) \right]$, and $f_i = f(E_{i\bm{k}})$ is the Fermi-Dirac distribution function evaluated at the quasiparticle energy $E_{i\bm{k}}(T)$,
and $\langle \cdots \rangle_{\mathrm{FS},i}$ represents an average over the $i$th Fermi-surface sheet.

\subsection*{Directional tunneling conductance}
\label{sec:STM}
We calculate the tunneling spectra of V$_2$Ga$_5$ using the three-gap anisotropic superconducting model with a fully symmetric singlet pairing.
The tunneling conductance is modeled within the tunneling-Hamiltonian formalism, assuming a normal-metal tip with a constant density of states. 
To the lowest order in the tunneling matrix elements, the differential conductance can be written as
\begin{equation}
\frac{dI}{dV}(V) \propto
\int dE
\left(-\frac{\partial f(E-eV)}{\partial E}\right)
N_{\mathrm{s}}(E),
\end{equation}
where $f(E)$ is the Fermi-Dirac distribution. For a multigap anisotropic superconductor, the superconducting density of states in Dynes formulation \cite{Dynes1978:PRL,Herman2016:PRB} entering the tunneling current is expressed as
\begin{equation}
N_{\mathrm{s}}(E, \hat{\bm n}) =
\sum_i
\left\langle
|T_{i\bm{k}}|^2\,
\mathrm{Re}
\left[
\frac{E - i\Gamma}
{\sqrt{(E - i\Gamma)^2 - \Delta_{i\bm{k}}^2(T)}}
\right]
\right\rangle_{\mathrm{FS},i},
\end{equation}
where $\Gamma$ is a phenomenological quasiparticle-broadening parameter.
Directional tunneling effects are incorporated through the tunneling matrix elements $|T_{i\bm{k}}|^2=v_{\rm g} D(\bm{k})$, where $v_{\rm g}$ is the group velocity at $i$th Fermi sheet defined as $v_{{\rm g}}=|\nabla_{\bm{k}} \xi_{i\bm{k}}\cdot\hat{\bm{n}}|$, and $\hat{\bm{n}}$ is the tunneling surface normal direction.
The directionality function $D(\bm{k})$ \cite{Ledvij1995:PRB,Yusof1998:PRB}, is given by
$D(\bm{k})=\exp(-\frac{\bm{k}^2-(\bm{k}\cdot\hat{\bm{n}})^2}{(\bm{k}\cdot\hat{\bm{n}})^2 \theta^2})$, where $\theta$ is an angular spread parameter controling tip size with an angular spread in $k$-space of the quasiparticle momenta with non-negligible tunneling probability with respect to $\hat{\bm{n}}$.

\subsection*{Multiparameter fitting}
The model parameters were determined by a joint fit to the electronic heat capacity and the directional tunneling conductance.
For each trial parameter set \({\cal P}\), containing the independent elements $V_{ij}$ of the coupling matrix and the gap-anisotropy parameters \(a\) and \(b\), the self-consistent multiband gap equation was solved first.
The resulting gap functions were then used without further adjustment to calculate both the heat capacity and the tunneling conductance for the two measured orientations, \((\hat{\bm{n}}\parallel [001])\) and \((\hat{\bm{n}}\perp [001])\).
In this way, the thermodynamic and spectroscopic data constrained the same superconducting gap structure.
The conductance spectra were calculated at the experimental temperature \(T = 0.52\)~K and normalized in the same high-bias region as the experimental spectra.
The optimization was performed by minimizing the weighted objective function
\begin{align}
\mathcal{J}({\cal P}) &=
\frac{1}{N_C}
\sum_{i=1}^{N_C}
\left[
C_{\cal P}(T_i)-C_i^{\mathrm{exp}}
\right]^2 \nonumber \\
&+
w_G
\sum_{\nu}
\frac{1}{N_\nu}
\sum_{j=1}^{N_\nu}
\left[
G_{\cal P}^{\nu}(V_j)-G_j^{\nu,\mathrm{exp}}
\right]^2 ,
\label{eq:joint_objective}
\end{align}
where \(C_i^{\mathrm{exp}}\) is the measured heat-capacity data, \(G_j^{\nu,\mathrm{exp}}\) is the measured tunneling conductance for orientation \(\nu\in\{\hat{\bm{n}}\parallel c,\hat{\bm{n}}\perp c\}\), and \(N_C\) and \(N_\nu\) are the corresponding numbers of fitted data points.
The weight \(w_G\) controls the relative contribution of the tunneling-conductance data with respect to the heat-capacity data; in the fits reported here, \(w_G = 0.8\).
The minimization was carried out using Optuna~\cite{akiba2019optuna}.

\subsection*{Data availability}
All data needed to evaluate the conclusions in the paper are available within the article. All raw data generated during the current study are available from the corresponding author upon request.

\subsection*{Code availability}
The postprocessing codes used in this study are available from the corresponding authors upon request.

\bibliographystyle{apsrev4-2}
\bibliography{V2Ga5}

\acknowledgments
This work was supported by the Slovak Research and Development Agency under the contract No. APVV-23-0624 and by the Ministry of Education, Research, Development and Youth of the Slovak Republic, provided under Grant No. VEGA 2/0073/24.
Research performed at Gdansk University of Technology was supported by the National Science Center (Poland), Project No. 2022/45/B/ST5/03916.
M.G.~acknowledges financial support provided by the Ministry of Education, Research, Development and Youth of the Slovak Republic, provided under Grant No. VEGA 1/0104/25 and the Slovak Academy of Sciences project IMPULZ IM-2021-42.

\subsection*{Author contributions}
J.H. developed the theoretical three-band model, implemented the numerical code, performed the self-consistent gap calculations, fitted the model to the experimental data, and drafted the manuscript.
J.K. conducted the specific heat and upper-critical-field measurements.
F.K., L.F. and P.Sz. carried out the tunneling spectroscopy measurements, and F.K. performed the upper-critical-field analysis.
Sz.K. and M.J.W. synthesized the single crystals and performed structural characterization.
T.K. supervised synthesis and characterization and acquired the funding.
P.S. contributed to the project conceptualization, supervised the experimental investigations, and participated in manuscript editing.
M.G. contributed to the project conceptualization, supervised the theoretical framework, performed the DFT electronic structure calculations, generated the Fermi surface maps, analyzed the theoretical results, and wrote the manuscript.
All authors participated in the scientific discussion of the results, provided comments, and approved the final manuscript.

\subsection*{Competing interests}
The authors declare no competing interests.


\clearpage 

\onecolumngrid          
\pagestyle{empty}       
\hoffset=-1in           
\voffset=-2in           
\topmargin=0pt          
\oddsidemargin=0pt      
\evensidemargin=0pt     

\foreach \x in {1,...,6} {
    \clearpage
    \thispagestyle{empty} 
    \noindent
    \vbox to \paperheight {
        \vss
        \hbox to \paperwidth {
            \hss
            \includegraphics[page=\x,width=\paperwidth,height=\paperheight,keepaspectratio]{supplementary_information.pdf}
            \hss
        }
        \vss
    }
}

\end{document}